\documentstyle[12pt]{article}

\begin{document}
\baselineskip 0.7cm

\newcommand{\gsim}{ \mathop{}_{\textstyle \sim}^{\textstyle >} }
\newcommand{\lsim}{ \mathop{}_{\textstyle \sim}^{\textstyle <} }
\newcommand{\EV}{ {\rm eV} }
\newcommand{\KEV}{ {\rm keV} }
\newcommand{\MEV}{ {\rm MeV} }
\newcommand{\GEV}{ {\rm GeV} }
\newcommand{\TEV}{ {\rm TeV} }
\newcommand{\ov}{ \over }
\renewcommand{\thefootnote}{\fnsymbol{footnote}}
\setcounter{footnote}{1}

\begin{titlepage}
\begin{flushright}
UT-834
\\
January, 1999
\end{flushright}

\vskip 0.35cm
\begin{center}
{\large \bf 
A Gauge-Mediation Model of Dynamical SUSY Breaking\\ 
with a Wide Range of the Gravitino Mass
}
\vskip 1.2cm
Izawa K.-I., Yasunori Nomura and T.~Yanagida
\vskip 0.4cm

{\it Department of Physics, University of Tokyo,\\
     Tokyo 113-0033, Japan}

\vskip 1.5cm

\abstract{
 We provide a gauge-mediation model of dynamical supersymmetry breaking
 where the gravitino mass takes a value in a wide range from $0.1~\EV$
 to $10~\GEV$.
 The lower mass region of order $100~\KEV$ or less deserves experimental
 and cosmological interests.
 The vacuum of our model is a true one without runaway instability,
 which may be desirable in a cosmological evolution of the universe. 
}
\end{center}
\end{titlepage}

\renewcommand{\thefootnote}{\arabic{footnote}}
\setcounter{footnote}{0}

%
%
%
%

\section{Introduction}

If it is caused by nonperturbative dynamics 
of some gauge interaction,
low-energy supersymmetry (SUSY) breaking
may provide a solution to the hierarchy problem.
In recent years, many mechanisms 
for dynamical SUSY breaking (DSB) have been found and 
many realistic gauge-mediation models have been 
constructed where the DSB effects are 
transfered to the standard-model sector through the known 
gauge interactions \cite{DNS}.
The gauge-mediation models of DSB are very attractive not only 
because they provide calculable models for the mass spectra 
of superparticles in the SUSY standard model (SSM), but also 
because they can naturally solve the flavor-changing neutral current 
problem and the CP violation problem in the SSM.

Most of the gauge-mediation models considered so far \cite{DNS} 
predict a gravitino mass $m_{3/2} \gsim 1~\MEV$ 
and few models are known to have $m_{3/2} \lsim 100~\KEV$
\cite{Nom, Nel}.
However, models with the lighter gravitino $m_{3/2} \lsim 100~\KEV$ 
are interesting experimentally since the lightest superparticle 
beside the gravitino can decay into the gravitino emitting photons
inside the detector, which may be testable in future 
experiments \cite{Exp}.
Also it has been pointed out \cite{Kaw}
that the lighter gravitino is 
cosmologically favored if string moduli exist at the electroweak 
scale.

The purpose of this paper is to provide a novel gauge-mediation
model where the gravitino mass takes a value in a wide range
$0.1~\EV \lsim m_{3/2} \lsim 10~\GEV$.%
\footnote{The upperbound of the gravitino mass $m_{3/2} \lsim 10~\GEV$
is determined so that the flavor-changing neutral currents induced by
superparticle loops are sufficiently suppressed.}
The vacuum of our model is a true one without runaway instability
\cite{Iza, Lut},
which may be desirable in a cosmological evolution of the universe.

\section{The Model}

Our model is based on two sectors of simple gauge theories.
One is an SU(2) gauge theory with 4 doublet 
chiral superfields $Q_i$,
where $i$ denotes a flavor index ($i = 1, \cdots, 4$).
The other is an SO(10) gauge theory with chiral superfields $H$
in the 10-dimensional representation and $\psi$ in the 16-dimensional
representation.  
The model consists of an SU(2) and an SO(10)
DSB models \cite{Yan, Mur}
combined by means of gauge singlets through a superpotential.

The superpotential is given by
\begin{equation}
 W = \lambda_Z Z_a (QQ)_a + \lambda_Y Y (QQ) + {1 \over 2}f_Y S Y^2
     + {1 \over 2}f_H S H^2 + k S q{\bar q},
\label{TSP}
\end{equation}
where $(QQ)_a$ ($a = 1, \cdots, 5$) and $(QQ)$ denote
suitable linear combinations \cite{Nom, Hot}
of the gauge invariants $Q_iQ_j$ and
$Z_a$, $Y$, $S$, $q$ and ${\bar q}$ are all singlets of
SU(2)$\times$SO(10).%
\footnote{Any symmetry which allows all the terms in Eq.~(\ref{TSP})
cannot ensure the absence of a term ${1 \over 2}f_Z S Z_a^2$ which
destabilizes the SUSY-breaking vaccum, so that we have to forbid this
term by hand.
However, the superpotential can be made natural if we take the
couplings between singlets and $Q$'s to be 
$\lambda_Z Z_a (QQ)_a + \lambda_Y Y (QQ)^2 / M^2$, where $M$ is the
gravitational scale.
Then, the gauge group acting on $Q$'s should be SP($2N$) ($N>1$, 
SP(2) $\simeq$ SU(2)) instead of SU(2), and 
$\Lambda^2$ should be replaced by $\Lambda^4 / M^2$ in the
following discussions \cite{Ken}.} 
It is straightforward to introduce messenger leptons in addition to
the messenger quarks $q$ and ${\bar q}$ \cite{DNS},
but here we omit them for simplicity of presentation.
The term $k S q{\bar q}$ transmits the DSB effects 
to the messenger quarks, which is a minimal setting of
gauge mediation and predicts a definite mass spectrum of superparticles
in the SSM \cite{DNS}.

In the following sections, we analyze the model by means of effective
field theories.
We adopt the coupling constants
of order one and a phase convention that they are positive.

\section{The Effective Superpotential}

We first integrate out the sector of the SU(2) gauge theory.
Assuming that the coupling $\lambda_Z$ is relatively large, we can also
integrate out the singlets $Z_a$ along with the doublets $Q_i$
to obtain \cite{Nom, Hot}
\begin{equation}
 W_{eff} = \lambda_Y \Lambda^2 Y + {1 \over 2}f_Y S Y^2
     + {1 \over 2}f_H S H^2 + k S q{\bar q},
\end{equation}
where $\Lambda$ denotes a dynamical scale of the SU(2) gauge interaction.

Without the SO(10) gauge interaction, the model would possess
a SUSY-invariant vacuum with $S \rightarrow \infty$.
This runaway behavior is circumvented by the SO(10) gauge
interaction, which gives SUSY-breaking effects
when the 10-dimensional
representation $H$ is decoupled by a mass $|f_H S|$ and
only the 16-dimensional representation $\psi$ is left over
\cite{Lut, Mur}.
Conversely, if the SU(2) gauge interaction were absent, the model would
have a SUSY-invariant vacuum with a mass $|f_H S|$ of the 10-dimensional
representation $H$ equal to zero \cite{Mur}.
In contrast to the models in Ref.~\cite{Iza, Lut},
the present model avoids an undesirable vacuum with $S=0$ due to 
the sector of the SU(2) gauge theory (see the next section).

The contribution of the SO(10) sector to the effective superpotential
is added \cite{Mur} to yield
\begin{equation}
 W_{eff} = \lambda_Y \Lambda^2 Y + {1 \over 2}f_Y S Y^2
     + {1 \over 2}f_H S H^2 + k S q{\bar q}
     + {2^{1 \ov 5} {\tilde \Lambda}^{21 \ov 5} \over (\psi \psi H)^{2 \ov 5}},
\end{equation}
where $\tilde \Lambda$ denotes a dynamical scale of the SO(10)
gauge interaction.%
\footnote{The SO(10) sector can be replaced by an SU(5)
DSB model \cite{Mur} without affecting our conclusions.}
We may write this effective superpotential as
\begin{eqnarray}
 W_{eff} = \lambda_Y \Lambda^2 Y + {1 \over 2}f_Y S Y^2
     + f_H S H^+ H^- + k S q{\bar q}
     + {{\tilde \Lambda}^{21 \ov 5} \over (\chi^2 H^+)^{2 \ov 5}}
\end{eqnarray}
in terms of variables $H_+$, $H_-$, $\chi$ defined in
Ref.~\cite{Mur}
with a $D$-flatness condition
\begin{equation}
 |H^+|^2 -|H^-|^2 - {1 \over 2}|\chi|^2 = 0.
\end{equation}
Here and henceforth, we denote the scalar component
of a chiral superfield by the same symbol as the corresponding superfield.

\section{The Effective Potential}

In the following analysis, we concern ourselves with the regime
${\tilde \Lambda} \lsim \sqrt{\lambda_Y} \Lambda$ of our interest.

\subsection{The $|f_H S| \ll {\tilde \Lambda}$ region}

Let the effective K{\"a}hler potential around the
minimum of the potential under the fixed value of $S$
be approximately canonical in the variables $H_+$, $H_-$, $\chi$
and the singlets in the effective theory.
This treatment seems adequate for small $|f_H S|$,
which results in large $|H|$
compared to ${\tilde \Lambda}$
\cite{Mur}.

The effective potential around the minimum under the fixed $S$ is given by
\begin{eqnarray}
 V_{eff} &\simeq& |\lambda_Y \Lambda^2 + f_Y S Y|^2
     + \left|{1 \over 2}f_Y Y^2 + f_H H^+ H^- + k q{\bar q}\right|^2
     + |kSq|^2 + |kS{\bar q}|^2
\nonumber \\
     & & + |f_H S H^+|^2 + \left|f_H S H^-
     - {2 \over 5}{{\tilde \Lambda}^{21 \ov 5} \over (\chi^2 H^+)^{7 \ov 5}}
     \chi^2 \right|^2
     + \left|{4 \over 5}
     {{\tilde \Lambda}^{21 \ov 5} \over (\chi^2 H^+)^{7 \ov 5}}
     \chi H^+\right|^2
\\
         &=& |\lambda_Y \Lambda^2 + Y'|^2
     + |q'|^2 + |{\bar q}'|^2
     + {1 \over |S|^4} \left|{1 \over 2}f_Y^{-1} Y'^2 + f_H^{-1} H'^+ H'^-
     + k^{-1} q' {\bar q}'\right|^2
\nonumber \\
     & & + |H'^+|^2 + \left|H'^-
     - {2 \over 5}{(f_H S)^{11 \ov 5}
     {\tilde \Lambda}^{21 \ov 5} \over (\chi'^2 H'^+)^{7 \ov 5}}
     \chi'^2 \right|^2
     + \left|{4 \over 5}{(f_H S)^{11 \ov 5} {\tilde \Lambda}^{21 \ov 5}
     \over (\chi'^2 H'^+)^{7 \ov 5}}
     \chi' H'^+\right|^2
\end{eqnarray}
with the $D$-flatness condition $|H'^+|^2 - |H'^-|^2 - |\chi'|^2/2 = 0$,
where the primed variables are defined by
\begin{eqnarray}
 Y' = f_Y S Y, \quad q' = k S q, \quad {\bar q}' = k S {\bar q},
 \quad H'^\pm = f_H S H^\pm, \quad \chi' = f_H S \chi.
\end{eqnarray}

When $k$ is relatively large, $q={\bar q}=0$ is energetically favored.
Then the effective potential around the
minimum under the fixed $S$
may be approximated by
\begin{eqnarray}
 V_{eff} \simeq |\lambda_Y \Lambda^2 + Y'|^2
     + {1 \over |S|^4}\left|{1 \over 2}f_Y^{-1} |Y'|^2 
     - f_H^{-1} |H'^-|^2\right|^2
     + 2|H'^-|^2
\end{eqnarray}
with the aid of $|H'^+|^2 = |H'^-|^2 + |\chi'|^2/2$, since 
$|f_H S| \ll {\tilde \Lambda} \lsim \sqrt{\lambda_Y} \Lambda$.
Hence the minimum of the potential with the fixed $S$ is given by
\begin{eqnarray}
 V_{eff} \simeq {f_H \lambda_Y^2 \Lambda^4 \over f_H + f_Y} - f_H^2 |S|^4,
\label{potential_1}
\end{eqnarray}
where the $F$ term of $S$ turns out to be
\begin{eqnarray}
 |F_S| \simeq f_H |S|^2.
\end{eqnarray}
The potential Eq.~(\ref{potential_1}) shows that $|S|$ tends to run away 
from the origin.

\subsection{The $|f_H S| \gg \tilde \Lambda$ region}

When $|f_H S|$ is large,
we may integrate out the sector of the
SO(10) gauge theory to obtain an effective potential of the form
\begin{eqnarray}
 V_{eff} &\simeq& |\lambda_Y \Lambda^2 + f_Y S Y|^2
     + \left|{1 \over 2}f_Y Y^2 + k q{\bar q}\right|^2 
     + |kSq|^2 + |kS{\bar q}|^2
     + c|f_H S|^{2 \over 11} {\tilde \Lambda}^{42 \over 11},
\end{eqnarray}
where $c$ is a positive constant of order one
\cite{Lut}.
When $k$ is relatively large,
$q={\bar q}=0$ is again energetically favored.
Then the effective potential is given by
\begin{eqnarray}
 V_{eff} &\simeq& |\lambda_Y \Lambda^2 + f_Y S Y|^2 
     + \left|{1 \over 2}f_Y Y^2\right|^2
     + c|f_H S|^{2 \over 11} {\tilde \Lambda}^{42 \over 11}.
\end{eqnarray}

For $|\sqrt{f_Y} S| \ll \sqrt{\lambda_Y} \Lambda$,
the minimum of the potential under the fixed $S$ is given by
\begin{eqnarray}
 V_{eff} \simeq \lambda_Y^2 \Lambda^4
         - {3 \over 2^{2 \ov 3}} \left( \lambda_Y \Lambda^2 
         \left| \sqrt{f_Y} S \right| \right)^{4 \ov 3}
         + c|f_H S|^{2 \over 11} {\tilde \Lambda}^{42 \over 11},
\end{eqnarray}
where the $F$ term of $S$ turns out to be
\begin{eqnarray}
 |F_S| \simeq {1 \over 2^{1 \ov 3}}
              \left( \lambda_Y \Lambda^2 
              \left| \sqrt{f_Y} S \right| \right)^{2 \ov 3}.
\end{eqnarray}

For $|\sqrt{f_Y} S| \gg \sqrt{\lambda_Y} \Lambda$,
the minimum of the potential under the fixed $S$ is given by
\begin{eqnarray}
 V_{eff} \simeq \left|{\lambda_Y^2 \Lambda^4 \over 2f_Y S^2}\right|^2
         + c|f_H S|^{2 \over 11} {\tilde \Lambda}^{42 \over 11},
\label{potential_2}
\end{eqnarray}
where the $F$ term of $S$ turns out to be
\begin{eqnarray}
 |F_S| \simeq \left|{\lambda_Y^2\Lambda^4 \over 2f_YS^2}\right|.
\end{eqnarray}
We now see that the runaway of $|S|$ observed above is stopped 
by the second term \cite{Lut} in the potential Eq.~(\ref{potential_2}).

\section{The Vacuum}

The analysis in the previous section shows
that the effective potential $V_{eff}$
has the true minimum at
\begin{eqnarray}
 |S| \simeq \left(11 \ov 2c\right)^{11 \ov 46}
            \left(\lambda_Y^2 \Lambda^4 \ov f_Y f_H^{1 \ov 11}
            {\tilde \Lambda}^{21 \ov 11}\right)^{11 \ov 23}
     \gg \sqrt{f_Y^{-1} \lambda_Y} \Lambda,
\end{eqnarray}
when ${\tilde \Lambda} \ll \sqrt{\lambda_Y} \Lambda$.
The gluino and gravitino masses
\cite{DNS}
are given by
\begin{eqnarray}
 m_{1/2} \simeq {\alpha_3 \ov 4\pi}\left|{F_S \ov S}\right|
         \simeq {\alpha_3 \ov 4\pi}
                \left|{\lambda_Y^2\Lambda^4 \over 2f_YS^3}\right|, \quad
 m_{3/2} \simeq \sqrt{{V_{eff} \ov 3M^2}},
\end{eqnarray}
where $\alpha_3=g_3^2/4\pi$ denotes the color gauge coupling, 
$V_{eff}$ is given by Eq.~(\ref{potential_2}) and $M$ is the
gravitational scale $\simeq 2.4 \times 10^{18}~\GEV$.
Since the couplings $\lambda_Y, f_Y, f_H$ and $c$ are of order one, 
the gluino and gravitino masses are given by 
$m_{1/2} \sim (\alpha_3 / 4\pi) 
({\tilde \Lambda} / \Lambda)^{63/23} \Lambda$ and 
$m_{3/2} \sim ({\tilde \Lambda} / \Lambda)^{42/23} 
(\Lambda^2 / \sqrt{3}M)$.
We certainly have a wide range of the gravitino mass by means of two
independent scales $\Lambda$ and ${\tilde \Lambda}$.
For instance, $m_{1/2} \simeq 200~\GEV$ and $m_{3/2} \simeq 100~\KEV$ are 
realized when $\Lambda \simeq 6 \times 10^8~\GEV$ 
and ${\tilde \Lambda} \simeq 10^7~\GEV$.

In the extreme case ${\tilde \Lambda} \sim \sqrt{\lambda_Y} \Lambda$,
the analysis in the previous section implies that
the gluino mass $m_{1/2}$ and gravitino mass $m_{3/2}$ are of order
$(\alpha_3 / 4\pi) {\tilde \Lambda}$
and of order ${\tilde \Lambda}^2 / \sqrt{3}M$,
respectively.
For the gluino mass $m_{1/2} \simeq 200~\GEV$ we get 
${\tilde \Lambda} \simeq 2 \times 10^4~\GEV$ and 
a very light gravitino $m_{3/2} \simeq 0.1~\EV$.

\section{Conclusion}

We have constructed a gauge-mediation
model of dynamical supersymmetry breaking
whose vacuum is a true one without runaway instability
\cite{Iza,Lut}.
Beside the SSM sector, the model consists of an SU(2) and an SO(10)
DSB models
\cite{Yan,Mur}
combined by means of gauge singlets through a superpotential
Eq.~(\ref{TSP}).

The gravitino mass is determined by the ratio of dynamical scales
of the SU(2) and SO(10) gauge interactions
and takes a value in a wide range
$0.1~\EV \lsim m_{3/2} \lsim 10~\GEV$.
In particular, a light mass of order $100~\KEV$ or less deserves 
experimental and cosmological interests.
The present model is unique in the point that it may accommodate 
$0.1~\EV \lsim m_{3/2} \lsim 1~\KEV$ with the simplest messenger term 
$k S q{\bar q}$ of gauge-mediated SUSY breaking.
It is well known \cite{PP} that this mass region for the gravitino has
no cosmological problem of gravitino overproduction.

\newpage

%
%
%
\newcommand{\Journal}[4]{{\sl #1} {\bf #2} {(#3)} {#4}}
\newcommand{\APJ}{Ap.~J.}
\newcommand{\CJP}{Can.~J.~Phys.}
\newcommand{\NC}{Nuovo Cimento}
\newcommand{\NP}{Nucl.~Phys.}
\newcommand{\PL}{Phys.~Lett.}
\newcommand{\PR}{Phys.~Rev.}
\newcommand{\PRep}{Phys.~Rep.}
\newcommand{\PRL}{Phys.~Rev.~Lett.}
\newcommand{\PTP}{Prog.~Theor.~Phys.}
\newcommand{\SJNP}{Sov.~J.~Nucl.~Phys.}
\newcommand{\ZP}{Z.~Phys.}

\end{document}